\documentclass[%
 reprint, 
superscriptaddress,
 amsmath,amssymb,
 aps,
 prl,
]{revtex4-1}
\usepackage{graphicx}
\usepackage{bm}
\usepackage{comment}

\usepackage{color}
\begin{document}

\preprint{APS/123-QED}

\title{Collective cell migration of epithelial cells driven by chiral torque generation}

\author{Takaki Yamamoto}
 \email{takaki.yamamoto@riken.jp}
\affiliation{%
Laboratory for Physical Biology, RIKEN Center for Biosystems Dynamics Research, Kobe 650-0047, Japan
}%
\author{Tetsuya Hiraiwa}
\affiliation{
Mechanobiology Institute, National University of Singapore, Singapore 117411, Singapore.
}
\affiliation{Universal Biology Institute, University of Tokyo, 7-3-1 Hongo Bunkyo-ku, Tokyo 113-0033, Japan}
\author{Tatsuo Shibata}
 \email{tatsuo.shibata@riken.jp}
\affiliation{%
Laboratory for Physical Biology,  RIKEN Center for Biosystems Dynamics Research, Kobe 650-0047, Japan
}%

\date{\today}

\begin{abstract}
Various multicellular tissues show chiral morphology.
Experimental studies have shown this can originate from cell chirality. 
However, no theory has been proposed to connect the cellular chiral torque and multicellular chiral morphogenesis. 
We propose a model of confluent tissue dynamics with cellular chiral torque. 
We found that cells migrate unidirectionally under a gradient of cellular chiral torque. 
While the migration speed varies depending on the tissue's mechanical parameters, 
it is scaled solely by a structural order parameter for liquid-to-solid transition in confluent tissues.

\end{abstract}

\maketitle

The establishment of left-right (LR) asymmetry in tissues and organs is an intriguing event in development, which involves the coordinated activity of cells and molecules~\cite{mogilnerCytoskeletalChiralitySwirling2015a,inakiCellChiralityIts2016,naganathanActomyosindrivenLeftrightAsymmetry2016,wanCellChiralityEmergence2016}. Since proteins as the basic components of biological systems are chiral molecules, the breaking of LR symmetry can be a collective property orchestrated by these chiral molecular components~\cite{hirokawaLeftRightDeterminationInvolvement2009,lebretonMolecularOrganismalChirality2018}.

For the LR asymmetry in tissue morphogenesis,
it has been shown that in {\it Drosophila}, the embryonic hindgut twists unidirectionally~\cite{taniguchiChiralityPlanarCell2011}, and the male genitalia undergoes a clockwise rotation when looked from the apical side~\cite{suzanneCouplingApoptosisPatterning2010b,satoLeftRightAsymmetric2015}.
These are epithelial tissues, which consist of epithelial cells that adhere to each other through cell-cell junctions~\cite{guillotMechanicsEpithelialTissue2013}. 
During the hindgut twisting, the hindgut epithelial cells exhibit chiral properties in their shape and other properties~\cite{taniguchiChiralityPlanarCell2011}. 
In the genitalia rotation, the surrounding epithelial cells move collectively in the clockwise direction driven by LR asymmetric positional rearrangements of cells, which rotates the genital disk \cite{satoLeftRightAsymmetric2015,hiraiwaWavePropagationJunctional2017a}. 
Even more generally, such chiral cell rearrangements can induce unidirectional cell migration within an epithelial tissue \cite{satoCellChiralityInduces2015a}.
The above examples suggest that a collective behavior of cellular chirality gives rise to the tissue LR asymmetry.

At the single cell level,
several types of cells have been reported to exhibit chiral dynamics.
Examples include the extension of neurites in cultured nerve cell~\cite{tamadaAutonomousRightscrewRotation2010}, the chiral movement of \textit{C. elegans} cell cortex~\cite{naganathanActiveTorqueGeneration2014},
nuclear rotation in melanophores of zebrafish~\cite{yamanakaRotatingPigmentCells2015},
and 
actin cytoskeleton swirling
in human foreskin fibroblasts (HFFs) cultured on a 
micro pattern~\cite{teeCellularChiralityArising2015a}.
Remarkably, in zebrafish melanophores
and the HFFs on the micro pattern, 
the single cells can cell-autonomously generate chiral torques
in such a way that 
the apical side  (top side) of cells exhibits rotations 
with respect to
the basal side (bottom side) which adheres to substrates.
Although a variety of specific mechanisms 
generate chiral torque at the cellular level,
many of these chiral properties are  
governed by active chiral processes of actomyosin cytoskeleton~\cite{furthauerActiveChiralFluids2012a,furthauerActiveChiralProcesses2013,furthauerTaylorCouetteMotor2012}. 

Since actomyosin is a ubiquitous component of eukaryotic cells, it is natural to consider that chiral torque generation is not a feature restricted to the specific cells mentioned above. 
In particular, we consider a situation where chiral torque is generated by 
individual single epithelial cells in a tissue, 
motivated by the LR symmetry breaking of the epithelial tissues~\cite{taniguchiChiralityPlanarCell2011,suzanneCouplingApoptosisPatterning2010b,satoLeftRightAsymmetric2015},
and by the fact that single cells can generate chiral torque~\cite{tamadaAutonomousRightscrewRotation2010,naganathanActiveTorqueGeneration2014,yamanakaRotatingPigmentCells2015,teeCellularChiralityArising2015a}.
Based on the cell vertex model (CVM)
\cite{nagaiDynamicCellModel2001},
we propose a theoretical model of dynamics of an epithelial tissue with a chiral torque generated by individual single cells.
Then we investigate the tissue scale dynamical emergent properties. 

To model the dynamics of an epithelial tissue with chiral torques generated by individual single cells, we use a two dimensional (2D) CVM.
In the CVM, cells are described by polygons with vertices and edges.
The position of the $i$th vertex is represented by $\vec{r}_i$.
The force acting on each vertex is given by the derivative $-\partial E(\{\vec{r}_i\})/\partial \vec{r}_i$ of a potential function $E(\{\vec{r}_i \})$,
given by
\begin{eqnarray}
E(\{\vec{r}_i \}) &=& \frac{K}{2} \sum_{\alpha=1}^{N} (A_{\alpha}-A_0)^2
+ \frac{K_p}{2}\sum_{\alpha=1}^{N}(P_{\alpha}-P_0)^2\nonumber\\
&+& \sum_{\langle i,j\rangle} \Delta{\Lambda}_{ij}(t) {\ell}_{ij}.
\label{vertex_eq1}
\end{eqnarray}
Here, the first term on the right hand side describes the area elasticity with the area $A_{\alpha}$ of cell $\alpha$, the elastic modulus $K$, and the preferred area $A_0$.
The second term defines the perimeter elasticity with the perimeter length $P_{\alpha}$ of cell $\alpha$, the elastic modulus $K_p$ and the preferred perimeter length $P_0$.
We assume for simplicity that these cellular mechanical properties $K,K_p,A_0,P_0$ are spatially homogeneous.
$N$ is the total number of cells. 
The third term is introduced to explicitly describe the fluctuation in the line tension $\Delta\Lambda_{ij}(t)$.
Here, $\ell_{ij}$ is the length of a cell edge between $i$th and $j$th vertices.
We introduce a fluctuating tension $\Delta{\Lambda}_{i j}(t)$ as a colored Gaussian noise with $\langle\Delta\Lambda_{ij}(t)\rangle=0$ and
$\langle \Delta{\Lambda}_{i j}(t_1) \Delta{\Lambda}_{kl}(t_2) \rangle = \delta_{ik}\delta_{jl}\sigma^2 e^{-|t_1-t_2|/{\tau}}$.
Such time-correlated fluctuation of line tension is reported in \cite{curranMyosinIIControls2017}. 
We set $\tau=1$ in this letter for simplicity. 
Hereafter, we choose the units of length and forces such that the elastic modulus $K_p$ and the preferred cell area $A_0$ are unity, i.e., $K_p=1$ and $A_0=1$. 
With these units, $P_0$ is a so-called target shape index, which is a ratio between perimeter and square root of area~\cite{farhadifarInfluenceCellMechanics2007,biDensityindependentRigidityTransition2015}. 
If a single cell tends to be circular, $P_0=2\sqrt{\pi}\sim3.54$. 
We set $P_0=3.54$ and $K=10$ to avoid large cellular shape deformation unless otherwise noted. 

\begin{figure}[t]
 \begin{center}
  \includegraphics[width=85mm]{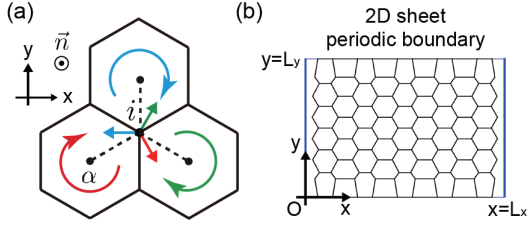}
 \end{center}
 \caption{(a) Schematic of the torque forces exerted on the $i$th vertex by the surrounding three cells generating chiral torque. The curved and solid arrows represent the chiral torque generation and torque forces, respectively, for positive $\nu_{\alpha}$. (b) 2D cellular sheet with the periodic boundary condition in the $x$-direction.}
 \label{setup}
\end{figure}

In the case of torque generation by isolated single cells, as reported in \cite{teeCellularChiralityArising2015a,yamanakaRotatingPigmentCells2015},
the apical surface rotates unidirectionally with respect to the basal substrate,
which is driven by the torque generated by actomyosin network.
In the epithelial tissue,
we also consider that a torque force is generated at the apical side in a specific direction with respect to the basal side, which adheres to the extracellular matrix (ECM).
To represent such a torque in the framework of the 2D CVM,
we introduce the torque force around the cell center as shown in Fig.~\ref{setup}(a). 
The simplest form may be given by
\begin{eqnarray}
\vec{T}_i=\sum_{\substack{{\rm cell}\ \alpha\\ {\rm around\ vertex}\ i}}\nu_\alpha(\vec{r}_i-\vec{r}_g^{\alpha})\times\vec{n},\label{torque_eq1}
\end{eqnarray}
where $\nu_{\alpha}$ is the coefficient of the torque force generated by the cell $\alpha$, $\vec{r}_g^{\alpha}$ is the area centroid of cell $\alpha$, 
and $\vec{n}$ is a unit normal vector from the basal to apical sides.

The time evolution equation for $\vec{r}_i$ is obtained by considering the force balance between
the frictional force $\eta{d\vec{r}_i}/{dt}$ with friction constant $\eta$,
the potential force $-\partial E(\{\vec{r}_i\})/\partial \vec{r}_i$ derived from Eq.(\ref{vertex_eq1}),
and the torque force $\vec{T}_i$ given by Eq.(\ref{torque_eq1}) as follows;
\begin{eqnarray}
\eta\cfrac{d\vec{r}_i}{dt}&=&-\cfrac{\partial E(\{\vec{r}_i\})}{\partial \vec{r}_i}  + \vec{T}_i.\label{timeevolution}
\end{eqnarray}
When the length of a cell edge falls below a threshold $l_{\rm th} = 0.03$ during a time-evolution according to Eq.(\ref{timeevolution}), a T1 transition is performed by flipping the edge by $90^\circ$
\cite{nagaiDynamicCellModel2001}.

We consider a simple configuration of a rectangular epithelial sheet with size $L_x$ and $L_y$ in the $x$- and $y$-direction, respectively, as shown in Fig.~\ref{setup}(b).
For the $x$-direction, we apply the periodic boundary condition at $x=0$ and $x=L_x$. 
The rectangular area size is equal to the number of cells so that the average cell area is set to be $\bar{A}_{\alpha}=1$.
This configuration corresponds to the tube geometry, a biologically ubiquitous structure of organs, which exhibits chiral morphogenesis as twisting of a heart tube and hindgut.
We simply consider that the cells are attached to the boundary. 
We hence fix the $y$-coordinates of vertices on the bottom and top boundaries to $y=0$ and $y=L_y$, respectively.
$\eta$ is set to $1$ unless otherwise noted.   
We calculate the time-evolution Eq.~(\ref{timeevolution}) with the time step $\Delta t =0.01$. 
We prepare initial cellular configurations packed with regular hexagonal cells to calculate the dynamics with $\sigma=0$ and $\nu_{\alpha}=0$ to relax the system, and then set $\sigma$ and $\nu_{\alpha}$ to the target values. 
The numbers of cells in the initial hexagonal configuration in the $x$- and $y$-directions are respectively denoted as $N_x$ and $N_y$, and we set $N_x=10$ for all the simulation. 
\begin{figure}[t]
 \begin{center}
  \includegraphics[width=85mm]{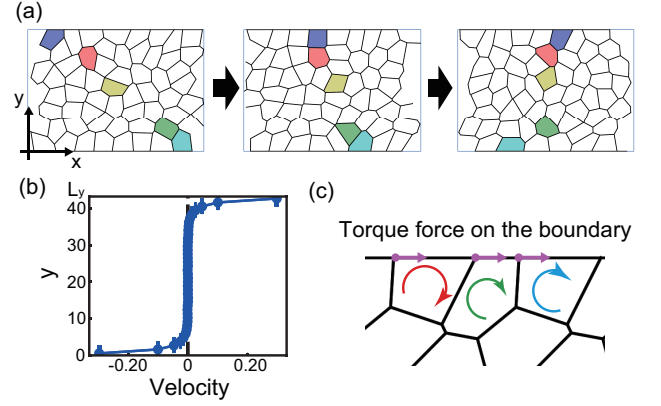}
 \end{center}
 \caption{(a) Time evolution of a cellular configuration in a numerical simulation ($\nu_{\alpha}=0.2, \sigma=0.3, N_y=6$). The time interval between each configuration is $40$ time units. Supplemental Material S1~\cite{supplemental} provides the movie. (b) Time-averaged flow profile for $N_y=40$ averaged over 4 samples ($\nu_{\alpha}=1.0, \sigma=0.3$). The error bars represent the standard deviation. (c) Schematic of the torque forces exerted on the vertices on a boundary ($\nu_{\alpha}>0$). The curved and solid arrows represent the chiral torque generation and the torque forces, respectively.}
 \label{homogeneous_torque}
\end{figure}

We first investigated the role of chiral torque generation on the dynamics of cells when the strength $\nu_{\alpha}$ of chiral torque generation is spatially homogeneous.
As shown in Fig.~\ref{homogeneous_torque}(a), we found that the cells migrate bidirectionally when the fluctuation in line tension is present ($\nu_{\alpha}=0.2, \sigma=0.3, N_y=6$). 
With negative $\nu_{\alpha}$, the cellular behavior was reversed, confirming that the chiral torque generation is the driving force of the cellular migration. 
For sufficiently large system ($N_y=40$), 
the flow profile rapidly decays near the boundary, and the bulk velocity vanishes (Fig.~\ref{homogeneous_torque}(b)). 
This indicates that the torque force and the potential force are balanced in the bulk, while not balanced near the boundary.  
On the top and bottom boundaries, the torque forces on the vertices are exerted in the rightward and leftward direction, respectively, as depicted in Fig.~\ref{homogeneous_torque}(c), leading to the bidirectional cellular flow at the boundaries. 

In the bulk, to understand how the torque force and the potential force are balanced, we consider a mean-field model where the torque force is exerted on a vertex $O$ surrounded by three regular hexagonal cells as shown in Fig.~\ref{setup}(a), without the line tension fluctuation.
When $\nu_{\alpha}$ is homogeneous, we readily find that the torque force exerted on the vertex $O$ vanishes due to the 3-fold symmetry. Therefore, even though the torque force is present in the bulk due to cellular deformation, it should be so small that the deformation of cells can readily balance it. 

We note that without the line tension fluctuation ($\sigma=0$),
all the cells are just deformed in the asymmetric fashion without continuous cellular flow (See Supplemental Material~\cite{supplemental}). 
Hence, the cell rearrangements induced by the stochastic fluctuation of the line tension promote the continuous cellular flow. 

\begin{figure}[t]
 \begin{center}
  \includegraphics[width=85mm]{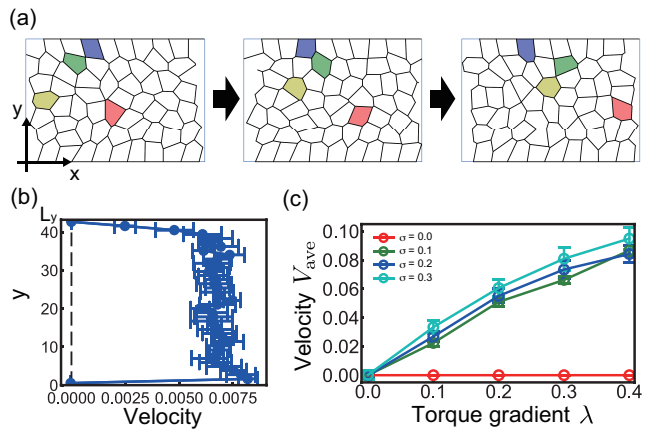}
 \end{center}
 \caption{(a) Time evolution of a cellular configuration in a numerical simulation ($\lambda=0.1, \sigma=0.3, N_y=6$). The time interval between each configuration is $40$ time units. Supplemental Material S2~\cite{supplemental} provides the movie. (b) Time-averaged flow profile for $N_y=40$ averaged over 5 samples ($\lambda=0.01, \sigma=0.3$). (c) Average velocity $V_{\rm ave}$ of cells is plotted against torque gradient $\lambda$ for different intensities $\sigma$ ($N_y=6$). The error bars represent the standard deviation.}
 \label{torque_gradient}
\end{figure}

We next consider a condition in which collective cell migration appears in a defined direction, induced by chiral torque generation. 
According to a symmetry argument~\cite{satoCellChiralityInduces2015a}, the symmetry along the $y$-axis has to be broken to achieve a unidirectional cellular movement along the $x$-axis.
In this letter, we consider a situation where chiral torque generation depends on the position of the cells along the $y$-axis.
We simply assume a linear form given by $\nu_\alpha	= -\lambda (y_g^\alpha - L_y)$,
where $y_g^{\alpha}$ is the $y$-coordinate of the center of cell $\alpha$.
Here, we impose a boundary condition in which the friction coefficients on the top and bottom boundaries are considerably higher ($\eta=10000$) to avoid the effect of the boundary torque force as we discussed above (Fig.~\ref{homogeneous_torque}(c)). 
With the torque gradient, we found that cells migrate unidirectionally in the direction perpendicular to that of the torque gradient (Fig.~\ref{torque_gradient}(a)). 
A time-averaged flow profile is shown in Fig.~\ref{torque_gradient}(b) ($\lambda=0.01,\sigma=0.3, N_y=40$).
With negative $\lambda$, the direction of the cellular migration is reversed. 
As $\lambda$ increases, the steady-state cellular velocity $V_{\rm ave}$ in the $x$-axis increases almost linearly (Fig.~\ref{torque_gradient}(c)). 
These results confirm that the chiral torque gradient is the driving force for this unidirectional collective cell migration. 
We also found that without the line tension fluctuation, the cells only deform without continuous flow (See Supplemental Material~\cite{supplemental}). 
Hence, the cell rearrangements induced by the stochastic fluctuation in the line tension are necessary for the continuous cellular flow.

\begin{figure}[t]
 \begin{center}
  \includegraphics[width=85mm]{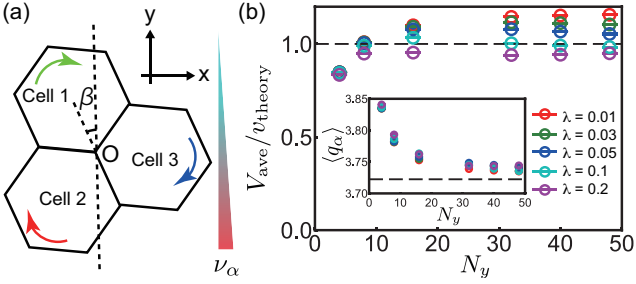}
 \end{center}
 \caption{(a) A cellular configuration for the theoretical analysis on the force exerted on a single vertex $O$. The curved arrows represent the direction of torque generations. (b) The normalized speed $V_{\rm ave}/v_{\rm theory}$ is plotted against the number $N_y$ of cells for different values of $\lambda$. The dashed line indicates $V_{\rm ave}/v_{\rm theory}=1$. 
 (Inset) The dependence of $\langle q_{\alpha}\rangle$ on $N_y$ is shown. The dashed line indicates $\langle q_{\alpha}\rangle=q_{\rm hex}$. }
 \label{fig_theory}
\end{figure}

In contrast to the case in which the strength of torque strength is homogeneous, 
the flow profile shown in Fig.~\ref{torque_gradient}(b) indicates that the cells can migrate in the bulk region under the torque gradient.
To see the mechanism of the unidirectional cellular migration, 
we consider the mean field model of three cells (Fig.~\ref{fig_theory}(a)). 
We set the position of the target vertex $O$ as the origin of the coordinate axes, and impose a linear torque gradient $\lambda$ along the $y$-direction.
Without loss of generality, we define the center of each surrounding cell as $\vec{r_g^{\alpha}}=l(\cos{(2\pi\alpha/3+\beta)}, \sin{(2\pi\alpha/3+\beta)})$, where $l=\sqrt{2\sqrt{3}}/3$ is the edge length for the hexagonal regular cell with area~$1$, $\beta$ is an arbitrary constant and the cell number $\alpha$ is indexed from $1$ to $3$.
In this configuration, since the force derived from the potential function $E$ vanishes in total owing to the 3-fold symmetry, only torque forces from the 3 cells contribute to the force exerted on the vertex $O$.
After a straightforward calculation, we obtain the force $\vec{f}=(3\lambda l^2/2, 0)$.
Consequently, the mean field model predicts that, when the cells are packed in a regular hexagonal pattern without any boundary, the cells migrate with the speed $v_{\rm theory}=3\lambda l^2/2\eta$.

We tested this prediction numerically using the model without fluctuation in line tension.
We apply a small value of $P_0=2.90$ to make the cell shape close to regular hexagon.
To suppress the boundary effect, we set the friction coefficients of the vertices on both of the bottom and top boundaries equal to those in the bulk, and set the torque force on the vertices on the boundaries to zero.
As shown in Fig.~\ref{fig_theory}(b), we found that, by increasing the system size $N_y$, the average cell speed $V_{\rm ave}$ approaches to $v_{\rm theory}$ for each $\lambda$, 
confirming that the spatial gradient of torque generated by the cells drives the cellular migration. 
The deviation from the theoretical value even for the large system size is owing to the inevitable cellular deformation from the regular hexagon assumed in the mean field model. 
In the inset of Fig.~\ref{fig_theory}(b), we confirmed the cellular deformation by calculating the average $\langle q_{\alpha}\rangle$ of the cell shape $q_{\alpha}= P_{\alpha}/\sqrt{A_{\alpha}}$, which equals to $q_{\rm hex}=2\sqrt{2}\sqrt[4]{3}\sim3.72$, when the cells are regular hexagon. 

\begin{figure}[t]
 \begin{center}
  \includegraphics[width=85mm]{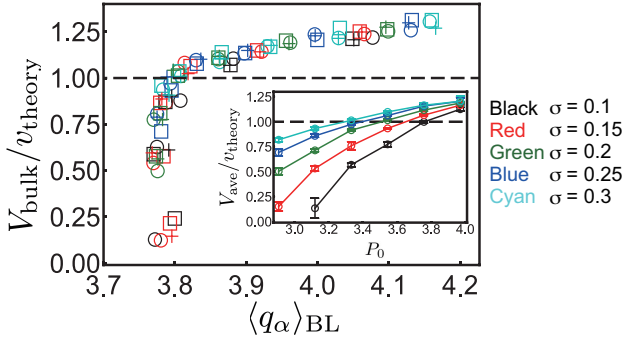}
 \end{center}
 \caption{The normalized bulk velocity $V_{\rm bulk}/v_{\rm theory}$ is plotted against the shape index $\langle q_{\alpha}\rangle_{\rm BL}$ in the boundary layers for different $\sigma$ and $P_0$ for a large system size $N_y=40$. 
 $\lambda$ is fixed to $0.01$. 
 The different symbols are for different samples. 
 (Inset) The normalized velocity $V_{\rm ave}/v_{\rm theory}$ is plotted against the target shape index $P_0$ for different values of $\sigma$. $V_{\rm ave}$ was averaged over 3 samples, and the error bars represent the standard deviation. }
 \label{velocity_vs_parameter}
\end{figure}

From the above analysis, we found that the cell shape affects the cellular migration velocity. 
The cell shape can be modulated by altering the target shape index $P_0$ or the noise strength $\sigma$. 
Therefore, we investigated how the migration speed $V_{\rm ave}$ depends on $P_0$ and $\sigma$ (inset of Fig.~\ref{velocity_vs_parameter}).
Here, we set $N_y=40,\lambda=0.01$ and applied the same boundary conditions that were used in Fig.~\ref{torque_gradient}.
We found that an increase of either $P_0$ or $\sigma$ enhances the migration speed (inset of Fig.~\ref{velocity_vs_parameter}). 
It has been reported that
an increase of $P_0$ turns the energy barrier for T1 transition to be lowered to almost zero above a critical value $P_0^*\sim3.81$, which is liquid-to-solid transition in confluent tissues~\cite{biDensityindependentRigidityTransition2015}.
Thus, as $P_0$ increases,
cell rearrangements occur more frequently.
The energy barrier is also overcome by increasing the line tension fluctuation $\sigma$, leading to an increase of the cell rearrangements frequency.
Hence, the migration velocity increases as either $P_0$ or $\sigma$ increases (inset of Fig.~\ref{velocity_vs_parameter}).

To see the influence of $P_0$ and $\sigma$ in a unified way,
we pay attention to the cell shape $q_{\alpha}$ obtained from the numerical simulations.
In Fig.~\ref{velocity_vs_parameter},
we plotted the migration velocity $V_{\rm bulk}$ in the bulk layers against the cell shape $\langle q_{\alpha}\rangle_{\rm BL}$ averaged in the boundary layers, and found that the velocity curves in the inset of Fig.~\ref{velocity_vs_parameter} are surprisingly collapsed in a single line.
Hence, $\langle q_{\alpha}\rangle_{\rm BL}$ is an excellent indicator of the migration velocity in our model.
Here, we define the boundary and bulk layers from the flow profile, and exclude the cells on the bottom and top boundaries for calculation of $\langle q_{\alpha}\rangle_{\rm BL}$, since the cells are strongly deformed by the flat boundaries (See Supplemental Material~\cite{supplemental}). 

We discuss how the bulk velocity~$V_{\rm bulk}$ is uniquely determined by the cell shape~$\langle q_{\alpha}\rangle_{\rm BL}$ in Fig.~\ref{velocity_vs_parameter}.
The cellular velocity should be determined by the balance between the bulk torque force and how easily cells rearrange over the energy barrier of T1 transition in the boundary layers. 
In the present model, since the torque force depends on the cell shape, the bulk torque force should be determined by the cell shape $\langle q_{\alpha}\rangle_{\rm BU}$ averaged in the bulk. 
The cell shape $q_{\alpha}$ has been also suggested as an indicator of how easily cells rearrange over the energy barrier of T1 transition~\cite{parkUnjammingCellShape2015a,biMotilityDrivenGlassJamming2016}.
Hence, the energy barrier of T1 transition in the boundary layers depends on $\langle q_{\alpha}\rangle_{\rm BL}$. 
Consequently, considering that $\langle q_{\alpha}\rangle_{\rm BL}\approx\langle q_{\alpha}\rangle_{\rm BU}$ is satisfied (See Supplemental Material~\cite{supplemental}), 
and hence both bulk torque force and energy barrier of T1 transition depend on $\langle q_{\alpha}\rangle_{\rm BL}$, 
$\langle q_{\alpha}\rangle_{\rm BL}$ uniquely determines the migration velocity.

In this letter, we have reported that a dynamical chiral property at the single-cell level, which is cellular chiral torque generation, can induce the collective cellular migration.
Our model predicts that, when the strength of torque is spatially homogeneous, the torque forces in the bulk almost balanced, and those generated by the cells at the boundaries drive the bidirectional cellular migration. 
Another prediction is that, under the gradient of chiral torque strength, the cells can migrate unidirectionally and perpendicularly to the gradient driven by the torque force generated in the bulk. 
Although the mechanism of the torque generation at the single-cell level is not fully revealed experimentally, previous studies reported that a combination of cytoskeleton and motor proteins generates the cellular torque. 
Activity of such cytoskeleton and motor proteins is regulated by various biochemical pathways such as Rho signaling pathway~\cite{etienne-mannevilleRhoGTPasesCell2002a}. 
Hence, the spatial gradient of regulatory molecules should produce a gradient of the strength of the cellular torque. 
In {\it in vivo} systems, an epithelial tissue is attached to other different tissues, and hence such attached tissues probably emit biochemical signals to generate the concentration gradient of molecules regulating the strength of the cellular torque. 
Since both the cellular chiral torque generation and the gradient of the cellular torque are considered to be ubiquitous in biological systems, we expect that the mechanism of the LR symmetry breaking in tissue dynamics we have proposed plays an essential role in the chiral collective cellular movements, such as a twist of epithelial tube~\cite{taniguchiChiralityPlanarCell2011} and a unidirectional epithelial cellular flow~\cite{satoLeftRightAsymmetric2015}, during development.
\begin{acknowledgements}
This work is supported by Grant-in-Aid for JSPS Fellows (Grant No. 18J01239 to TY),  KAKENHI Grant No. 17H07366 (to TY),  JP16K17777 (to TH), JP19K03764 (to TH) and JP19H00996 (to TS), and JST CREST grant number JPMJCR1852, Japan (TS).
\end{acknowledgements}

\end{document}


\title{SUPPLEMENTAL MATERIAL\\
Collective cell migration of epithelial cells driven by chiral torque generation}

\author{Takaki Yamamoto}
 \email{takaki.yamamoto@riken.jp}
\affiliation{%
Laboratory for Physical Biology, RIKEN Center for Biosystems Dynamics Research, Kobe 650-0047, Japan
}%
\author{Tetsuya Hiraiwa}
\affiliation{
Mechanobiology Institute, National University of Singapore, Singapore 117411, Singapore.
}
\affiliation{Universal Biology Institute, University of Tokyo, 7-3-1 Hongo Bunkyo-ku, Tokyo 113-0033, Japan}

\author{Tatsuo Shibata}
 \email{tatsuo.shibata@riken.jp}
\affiliation{%
Laboratory for Physical Biology,  RIKEN Center for Biosystems Dynamics Research, Kobe 650-0047, Japan
}%

\maketitle
\section{Definition of the area centroid of a cell}
We define the cell center of cell $\alpha$ using the area centroid as $\vec{r}_g^{\alpha}=\frac{1}{6A_{\alpha}}\sum_{\substack{\mu=1}}^{N_{\alpha}}(\vec{r}_{\mu}^{\alpha}+\vec{r}_{\mu+1}^{\alpha})(\vec{r}_{\mu}^{\alpha}\times\vec{r}_{\mu+1}^{\alpha})\cdot\vec{n}$. 
Here, $\mu$ is the index of the $N_{\alpha}$ vertices around the cell $\alpha$, indexed in the counterclockwise direction.
The position vector of each vertex is represented by $\vec{r}_{\mu}^{\alpha}$ and $\vec{r}_{N_{\alpha}+1}^{\alpha}=\vec{r}_{1}^{\alpha}$. 

\section{Preparation of the initial cellular configuration}
\begin{figure}[htbp]
\begin{centering}
\includegraphics[width = 15cm]{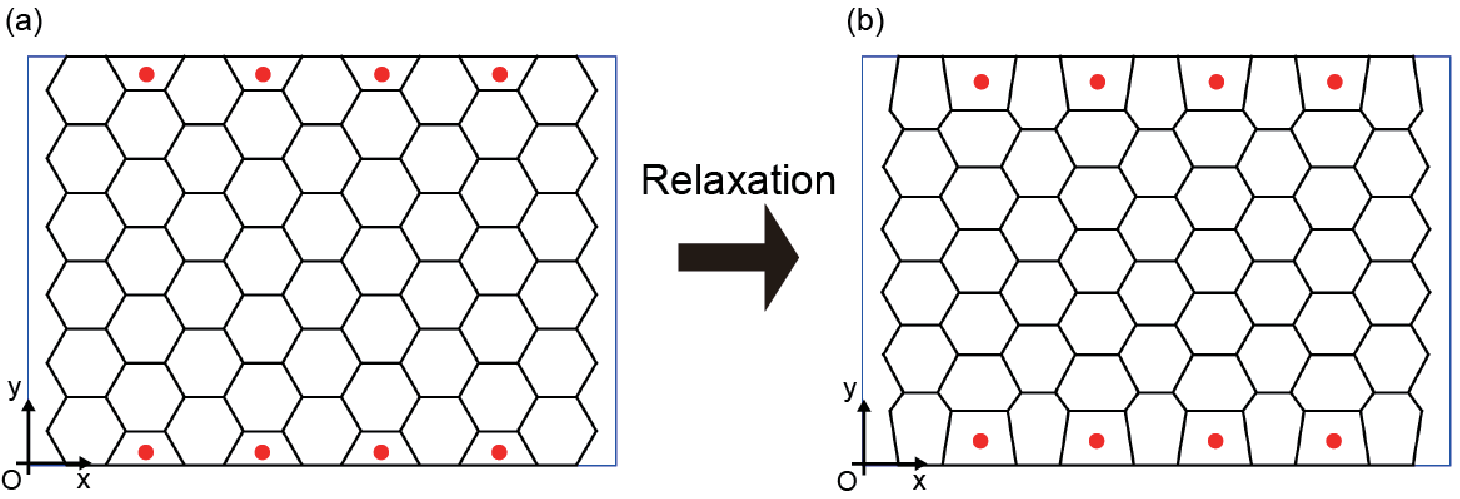}
\caption{\label{fig_s1}(a) The initial condition with regular hexagonal cells. The half regular hexagonal cells on the boundaries are shown with red dots. (b) A configuration after relaxation with $\sigma=0$ and $\nu_{\alpha}=0$ ($P_0=3.54$). }
\end{centering}
\end{figure}

We prepared the initial condition as shown in Fig.~\ref{fig_s1}(a). 
In the initial condition, the cells in the bulk are set to regular hexagonal shape, while the cells at the boundaries, which are labeled with red dots in Fig.~\ref{fig_s1}(a), are set to half of the regular hexagonal cells. 
The average cell area $\bar{A}$ is set to 1. 
Then, we calculate the dynamics with $\sigma=0$ and $\nu_{\alpha}=0$ to relax the system for 50 time units, and then we set $\sigma$ and $\nu_{\alpha}$ to the target values. 
Fig.~\ref{fig_s1}(b) is the configuration after the relaxation ($P_0=3.54$).

\section{Configuration obtained when $\sigma=0$. }

\begin{figure}[htbp]
\begin{centering}
\includegraphics[width = 12cm]{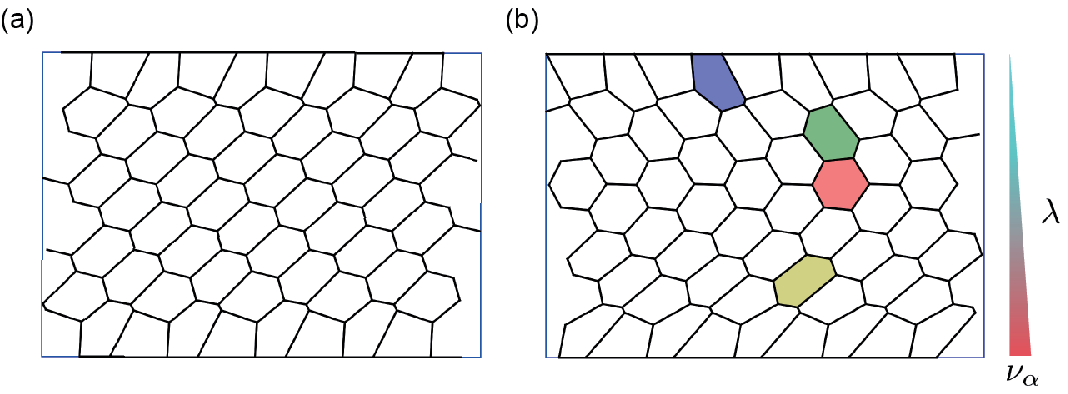}
\caption{\label{fig_s2}The tissue deformation observed without continuous cellular flow in the zero-noise limit $\sigma = 0$ (a) under a homogeneous cellular chiral torque ($\nu_{\alpha}= 0.2$) and (b) under a gradient of cellular chiral torque ($\lambda= 0.1$).}
\end{centering}
\end{figure}

\section{Definition of the boundary layers}
We systematically defined the boundary layers near either the bottom or the top boundaries. 
We discretized the area into $N_y$ layers from the $0$th layer to the $(N_y-1)$th layer. 
The $i$th layer is defined as the layer ranging from $y=iL_{y}/N_y$ to $y=(i+1)L_{y}/N_y$ ($i=0\sim(N_y-1)$). 
The boundary layers are defined as the $0$th$\sim$$N_{\rm b}$th and $N_{\rm t}$th$\sim$$(N_y-1)$th layers, respectively for those near the bottom and top boundaries. 

\begin{figure}[htbp]
\begin{centering}
\includegraphics[width = 12cm]{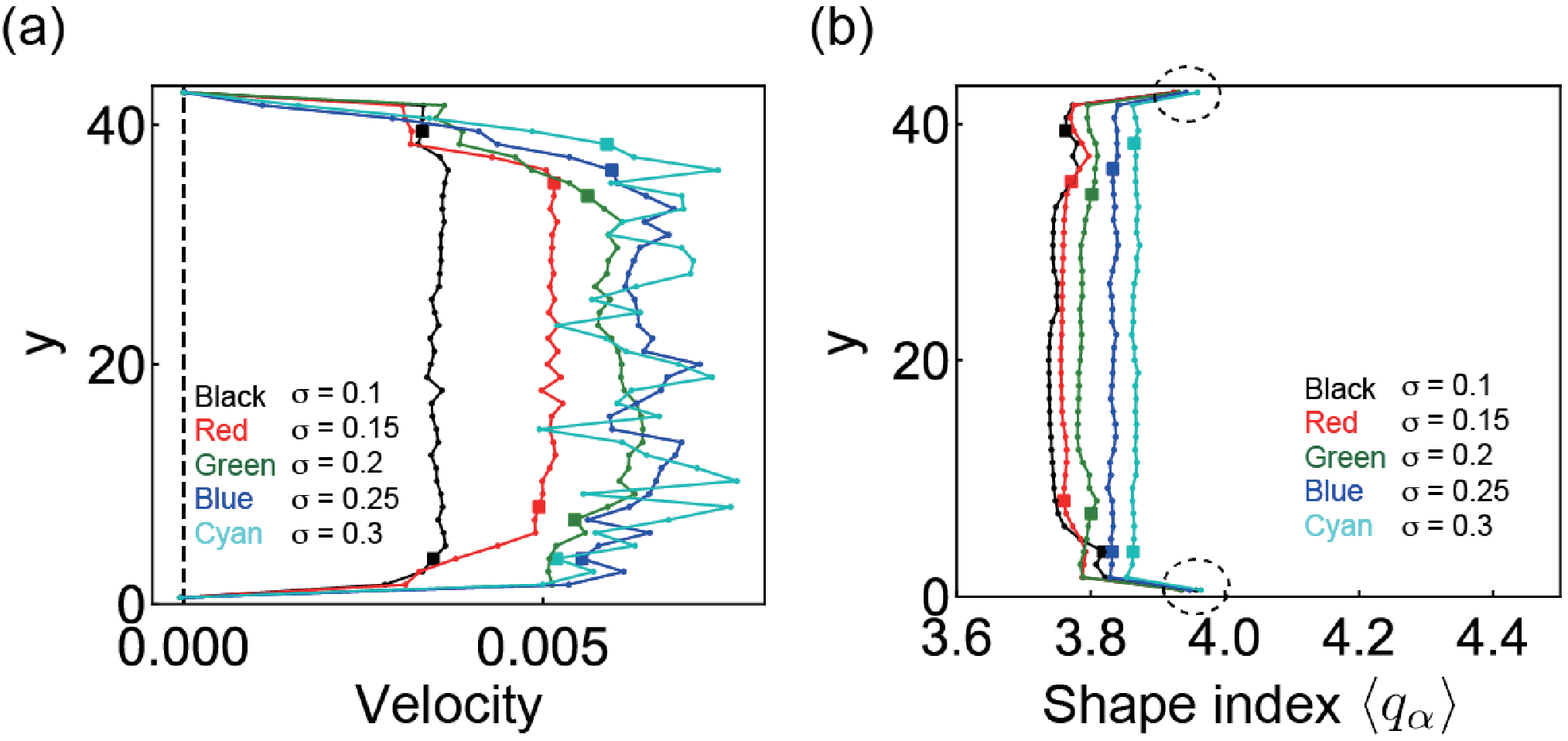}
\caption{\label{fig_s3}(a) Examples of the flow profiles are shown ($N_y=40,\lambda=0.01,p_0=3.33$). The $N_b$th and $N_t$th layers, determined by the above procedure, are also shown with the square markers. (b) Spatial profiles of shape index $\langle q_{\alpha}\rangle$ are shown. The $N_b$th and $N_t$th layers are also shown with the square markers. The data points for the top and bottom layers are marked with dashed ellipsoids as a guide for eye. In (a) and (b), each data point corresponds to each layer.}
\end{centering}
\end{figure}

We calculated the average velocity of each layer by averaging the velocity of cells in the layer in time, and drew the velocity profiles as shown in Fig.~\ref{fig_s3}(a) ($N_y=40,\lambda=0.01,P_0=3.33$). 
Using the velocity profile, we defined the boundary layers via the following procedure. 

\begin{enumerate}
\item We calculated the approximate bulk velocity $V_{\rm bulk}'$ and the approximate standard deviation $\sigma_{\rm bulk}'$ of the bulk velocity by averaging the velocity of the $15$th $\sim$ $24$th layers. 
\item We smoothed the velocity profile to obtain $v=v^{\rm sm}(i)$ by a simple moving average, where the mean is taken from two data on either side of a central value. 
\item We determined $N_{\rm b}$ as the maximum integer $i<15$ which satisfies either $v^{\rm sm}(i)<V_{\rm bulk}'-3\sigma_{\rm bulk}'$ or $v^{\rm sm}(i)>V_{\rm bulk}'+3\sigma_{\rm bulk}'$. Also, $N_{\rm t}$ is determined as the minimum integer $i>24$ which satisfies either $v^{\rm sm}(i)<V_{\rm bulk}'-3\sigma_{\rm bulk}'$ or $v^{\rm sm}(i)>V_{\rm bulk}'+3\sigma_{\rm bulk}'$.
\end{enumerate}

In Fig.~\ref{fig_s3}(a), we show the $N_b$th and $N_t$th layers, determined by the above procedure, with the square markers on examples of velocity profiles ($N_y=40,\lambda=0.01,P_0=3.33$). 

\section{Relationship between $\langle q_{\alpha}\rangle_{\rm BL}$ and $\langle q_{\alpha}\rangle_{\rm BU}$}

In Fig.~\ref{fig_s3}(b), we show examples of spatial profiles of the shape index $\langle q_{\alpha}\rangle$. 
We find that the shape index for the top and bottom layers are outlier due to the flat boundaries which induce large cell deformation. 
To avoid the outliers, we eliminated the top and bottom layers when we calculated $\langle q_{\alpha}\rangle_{\rm BL}$. 

\begin{figure}[htbp]
\begin{centering}
\includegraphics[width = 12cm]{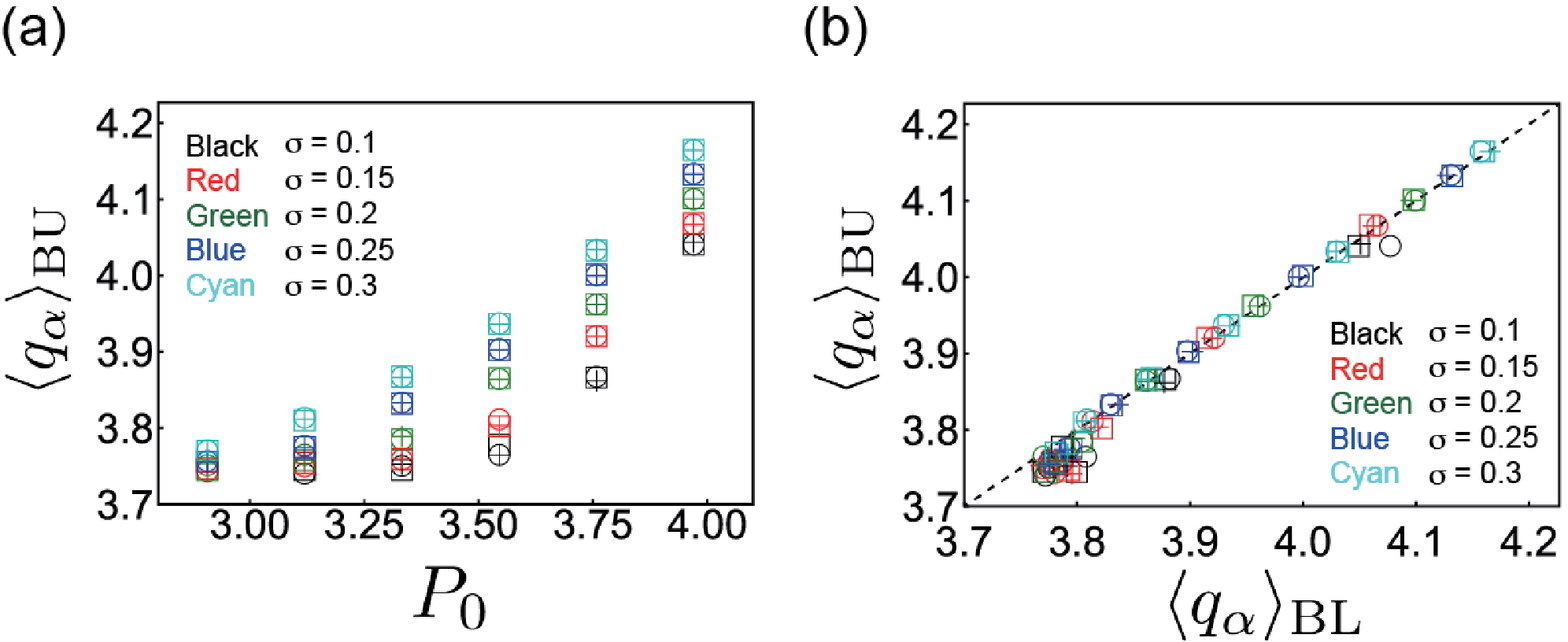}
\caption{\label{fig_s4}
(a) $\langle q_{\alpha}\rangle_{\rm BU}$ is plotted against $P_0$ for different values of $\sigma$ for a large system size $N_y=40$. The torque gradient $\lambda$ is fixed to $0.01$. The different symbols are for three different samples. (b) $\langle q_{\alpha}\rangle_{\rm BU}$ is plotted against $\langle q_{\alpha}\rangle_{\rm BL}$ for different $\sigma$ and $P_0$. 
 The dashed line indicates $\langle q_{\alpha}\rangle_{\rm BU}=\langle q_{\alpha}\rangle_{\rm BL}$. The different symbols are for three different samples. }
\end{centering}
\end{figure}

In Fig.~\ref{fig_s4}(a), we show the dependence of $\langle q_{\alpha}\rangle_{\rm BU}$ on the target shape index $p_0$ for the data in Fig.~5 in the main text ($N_y=40,\lambda=0.01,P_0=3.33$). 
We found that larger $\sigma$ and $p_0$ provide larger $\langle q_{\alpha}\rangle_{\rm BU}$. 

In Fig.~\ref{fig_s4}(b), we show the relationship between $\langle q_{\alpha}\rangle_{\rm BL}$ and $\langle q_{\alpha}\rangle_{\rm BU}$ for the same data set. 
We find $\langle q_{\alpha}\rangle_{\rm BU}\approx\langle q_{\alpha}\rangle_{\rm BL}$. 
As shown in Fig.~\ref{fig_s3}(b), the spatial profiles of the shape index are nearly homogeneous except on the top and bottom layers. 
Hence, we obtain $\langle q_{\alpha}\rangle_{\rm BU}\approx\langle q_{\alpha}\rangle_{\rm BL}$.

